\documentclass[journal,twoside,web]{ieeecolor}
\usepackage{tmi}
\usepackage{cite}
\usepackage{amsmath,amssymb,amsfonts}
\usepackage{arydshln}
\usepackage{multirow}
\usepackage[noend]{algpseudocode}
\usepackage{algorithmicx,algorithm}
\usepackage{graphicx}
\usepackage{textcomp}
\usepackage{epstopdf}
\usepackage{booktabs}
\usepackage[section]{placeins}
\usepackage{stfloats}
\usepackage{bm}
\usepackage{xstring}
\usepackage{bbding}
\usepackage{color}
\usepackage{ulem}
\definecolor{darkkcyan}{RGB}{0,138,218}
\usepackage[colorlinks=true, allcolors=darkkcyan,urlcolor = black]{hyperref}
\usepackage[switch]{lineno}
\useunder{\uline}{\ul}{}

\def\BibTeX{{\rm B\kern-.05em{\sc i\kern-.025em b}\kern-.08em
	T\kern-.1667em\lower.7ex\hbox{E}\kern-.125emX}}

\begin{document}
\title{3D Segmentation Guided Style-based Generative Adversarial Networks for PET Synthesis}
\author{Yang Zhou*, Zhiwen Yang*, Hui Zhang, Eric I-Chao Chang, Yubo Fan, and Yan Xu
	\thanks{This work is supported by the National Science and Technology Major Project of the Ministry of Science and Technology in China under Grant 2017YFC0110903, the National Natural Science Foundation in China under Grant 62022010, 81771910, SinoUnion Healthcare Inc. under the eHealth program, the Fundamental Research Funds for the Central Universities of China from the State Key Laboratory of Software Development Environment in Beihang University in China, the 111 Project in China under Grant B13003, the high performance computing (HPC) resources at Beihang University. \textit{(*Equal contribution. Corresponding author: Yan Xu.)} }
	\thanks{Y. Zhou, Z. Yang, Y. Fan, and Y. Xu are with the School of Biological Science and Medical Engineering, State Key Laboratory of Software Development Environment, Key Laboratory of Biomechanics and Mechanobiology of Ministry of Education, Research Institute of Beihang University in Shenzhen, Beijing Advanced Innovation Center for Biomedical Engineering, Beihang University, Beijing 100191, China (e-mail: ZhouYangBME@buaa.edu.cn; upyzwup@buaa.edu.cn; yubofan@buaa.edu.cn; xuyan04@gmail.com).}
	\thanks{H. Zhang is with the Department of Biomedical Engineering, Tsinghua University, Beijing 100084, China (e-mail: hzhang@tsinghua.edu.cn).}
	\thanks{E. I.-C. Chang is with Microsoft Research, Beijing 100080, China (e-mail: echang@microsoft.com).}}
\maketitle
\begin{abstract}
Potential radioactive hazards in full-dose positron emission tomography (PET) imaging remain a concern, whereas the quality of low-dose images is never desirable for clinical use. So it is of great interest to translate low-dose PET images into full-dose. Previous studies based on deep learning methods usually directly extract hierarchical features for reconstruction. We notice that the importance of each feature is different and they should be weighted dissimilarly so that tiny information can be captured by the neural network. Furthermore, the synthesis on some regions of interest is important in some applications. Here we propose a novel segmentation guided style-based generative adversarial network (SGSGAN) for PET synthesis. (1) We put forward a style-based generator employing style modulation, which specifically controls the hierarchical features in the translation process, to generate images with more realistic textures. (2) We adopt a task-driven strategy that couples a segmentation task with a generative adversarial network (GAN) framework to improve the translation performance. Extensive experiments show the superiority of our overall framework in PET synthesis, especially on those regions of interest.
\end{abstract}


\section{Introduction}
\label{sec:introduction}
\IEEEPARstart{P}{ositron} emission tomography (PET), an imaging modality that visualizes the metabolic process of the human body, has been widely used for disease diagnosis. After acquiring PET images, quantification of the regions of interest (ROIs) such as organs and lesions can be measured for clinical use. However, high-quality PET images require sufficient doses of radioactive tracers, which raises concerns about the risk of radiation exposure. Low-dose PET images, on the other hand, suffer from low signal-to-noise ratio (SNR). Thus, it is relevant to translate low-dose images into full-dose while remaining crucial detailed information.

A wide range of studies has been done for PET image-to-image translation (also called PET synthesis) from low-dose to full-dose. Traditional image processing methods such as regression forest\cite{Predictionofstandard-dosebrainPETimagebyusingMRIandlow-dosebrain}, sparse representation\cite{Predictingstandard-dosePETimagefromlow-dosePETandmultimodalMRimagesusingmapping-basedsparserepresentation}, and dictionary learning\cite{SemisupervisedTripledDictionaryLearningforStandard-DosePETImagePredictionUsingLow-DosePETandMultmodalMRI,Multi-levelcanonicalcorrelationanalysisforPETimageestimation} are applied to realize full-dose PET synthesis. However, these methods still have common flaws like time-consuming and over-smoothed generation effects. In the past few years, convolutional neural networks (CNNs) have been proved powerful in computer vision. In the particular case of PET synthesis, Xiang \textit{et al.} propose a deep auto-context CNN\cite{deepautocontext} for 2D PET slice estimation and it achieves satisfactory performance. Xu \textit{et al.} use a modified Unet\cite{200xLow-dosePETReconstructionusingDeepLearning} with skip connection, taking multiple slices as inputs (2.5D), to produce images preserving more local information. However, neither 2D or 2.5D data exploits spatial information well, resulting in a discontinuous estimation between slices of 3D image. To make better use of integrating context information, Wang \textit{et al.} propose 3D-cGAN\cite{wang20183d} taking 3D low-dose volumes as inputs, which has been reported to outperform 2D-cGAN\cite{mirza2014conditional} in all measures. Importantly, the use of generative adversarial networks (GAN)\cite{goodfellow2014generative} facilitates the generator to capture more in-depth information and to generate images with more detailed textures, mitigating the over-smoothed effect.

Recently, style modulation has become a popular approach for realistic image generation\cite{karras2019style, karras2020analyzing, park2019semantic}. Karras \textit{et al.} propose a novel generative model called StyleGAN \cite{karras2019style} for random portrait generation. It can produce a variety of high-resolution portraits by scale-specifically controlling styles and yield state-of-the-art performance in generative models. Inspired by this, we recognize that the gap between full-dose and low-dose images could be partly regarded as the differences in styles. When translating from low-dose to full-dose, the style of PET images has changed. So it is natural to think of modulating low-dose style to achieve full-dose performance in the translation process. 

Besides, clinicians mainly focus on pivotal ROIs, such as the liver and the kidney. So it is an important task to preserve these ROIs in the translation process. Researchers have shown the supremacy of task-driven-based generative models\cite{segmentationGuided, taskGAN}. They couple a GAN-based image generation model with another task-specific model to boost the generation performance in task-related regions. Therefore, task-driven strategy could be helpful to employ the segmentation task of PET to improve the image generation performance, especially in segmentation-task-related regions. Introducing the knowledge of segmentation into the training of the generation model can significantly enhance the synthesis image quality, especially in the segmentation-task-related regions.
\begin{figure*}[t]
	\centering
	\includegraphics[width=\textwidth]{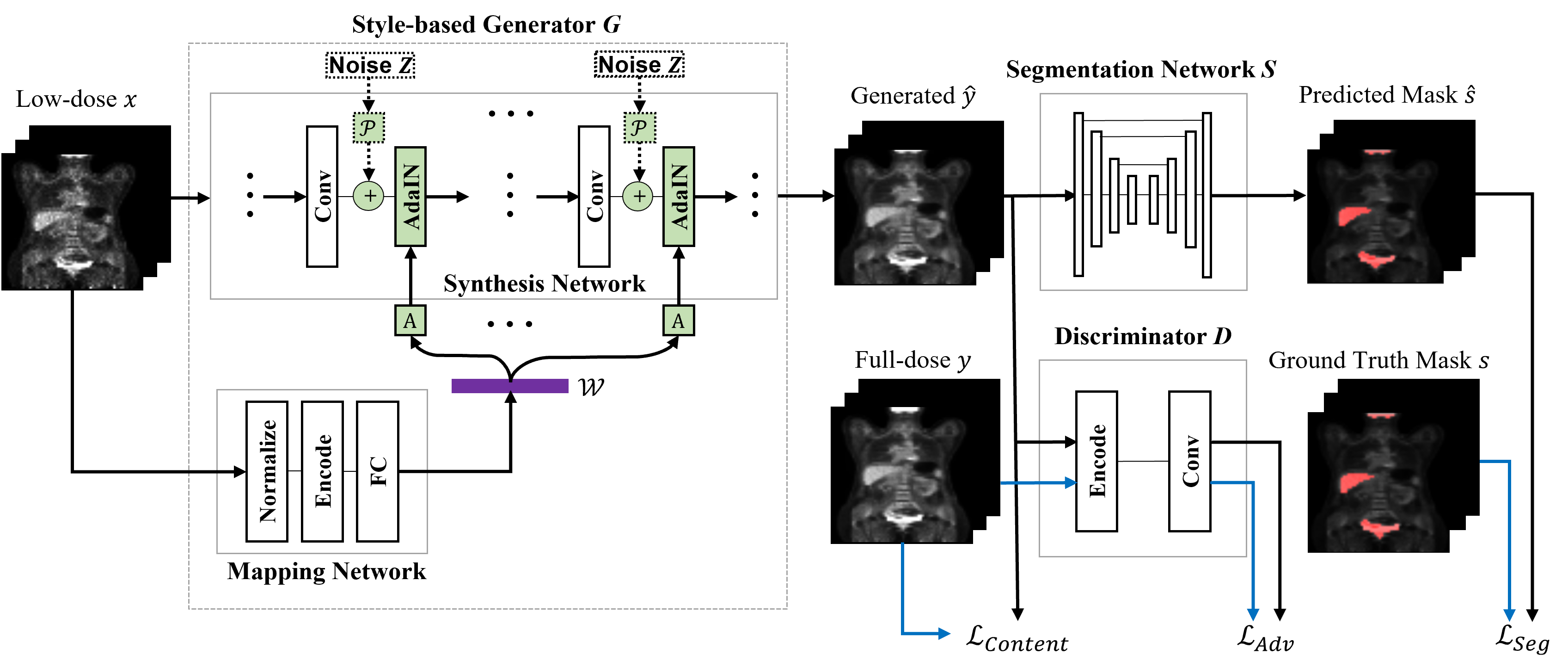}
	\caption{Framework of SGSGAN. SGSGAN includes a style-based generator $G$, a discriminator $D$, and a segmentation network $S$. The style-based generator $G$ consists of a mapping network that yields the style representation $\mathcal{W}$ and a synthesis network that conducts style modulation and image generation. $G$ is trained to synthesize 3D full-dose PET $\hat{y}$ from 3D low-dose PET $x$, while the discriminator $D$ is learned to distinguish between the generated image $\hat{y}$ and the real full-dose image $y$. The segmentation network $S$ is trained to back-propagate the segmentation loss to guide image generation. The objective function of SGSGAN is composed of three parts: $\mathcal{L}_{Adv}$, $\mathcal{L}_{Content}$ and $\mathcal{L}_{Seg}$. Note that the noise module is enabled in training and disabled in testing, which is marked with black dash lines.}
	\label{method}
\end{figure*}
\par In this paper, enlightened by the appealing success of the aforementioned ideas, we propose a segmentation guided style-based generative network (SGSGAN) for generating a full-dose PET image from a low-dose PET with segmentation as guidance. The overall framework of our method is shown in \autoref{method}. It consists of a style-based generator, a discriminator, and a segmentation network: (1) The style-based generator is where style modulation and PET translation are conducted. It takes a low-dose image as input to extract features hierarchically. We notice that the importance of each feature is different and they should be weighted dissimilarly. Style modulation well tackles this problem. After each convolutional layer, the output features will be rescaled and shifted accordingly using learned style factors for a more reasonable composition. So the generator can better learn to bridge the gap between low-dose and full-dose domains. (2) The discriminator conducts adversarial training with the generator and pushes the generator towards target distribution. (3) A segmentation guided strategy is employed to couple a segmentation task with the GAN framework for preserving and enhancing the denoised image quality in the segmentation-task-related ROIs of PET, which are the main concerns of clinicians. Specifically, a segmentation network is trained to produce the masks of generated images and back-propagate the segmentation loss to guide image generation. Details of our method will be described in \ref{sec_method}.

The novelties and contributions of this paper are summarized as follows:
\begin{itemize}
    \item A style-based generator is proposed for PET translation through style modulation.
    \item A segmentation task is introduced as guidance, namely the segmentation guided strategy, for better PET image synthesis, especially on those regions of interest.
	\item Our framework is flexible because the synthesis network could be implemented by varied network architectures (See \autoref{method}).
\end{itemize}

\section{Related Work}
\label{sec:related work}

Our work is closely related to three broad categories: (1) image-to-image translation, (2) style transfer, and (3) task-driven strategy.

\subsection{Image-to-image Translation}
Image-to-image translation is the kind of problem that synthesizes images from images, such as segmentation\cite{long2015fully}, edge detection\cite{xie2015holistically}, and image colorization\cite{iizuka2016let}. Pix2pix\cite{Isola2017image} is a representative model that applies conditional GAN (cGAN)\cite{mirza2014conditional} to image-to-image translation. Pix2pix is able to learn a high-quality mapping from an input distribution to a natural output distribution, but the results still lack structure details.  

Recently, a multiplicity of works on medical image restoration have adopted cGAN as their basic framework, including denoising\cite{yang2018low}, synthesis\cite{wang20183d}, and super-resolution\cite{chen2018brain,chen2018efficient}. But most of these existing methods still suffer from rather severe blurs and artifacts in synthesized images. MedGAN\cite{armanious2020medgan} and Ea-GAN\cite{yu2019ea} add additional perceptual constraints or gradient difference constraints to loss function in order to reduce artifacts and blurs of the generated images. However, they do not use style modulation to narrow the difference of style between the input image and the output image, causing discrepancies in style. CycleGAN has also been used for full-dose PET generation from low-dose PET\cite{supervisedCycleGAN,zhao2020study,sanaat2021deep,2019Whole}, but its unpaired architecture reduces the performance of synthesis compared with paired ones.

Our work focuses on 3D PET translation from low-dose images to full-dose images. We apply cGAN\cite{mirza2014conditional} as the basic framework of synthesis because this problem needs to establish a certain mapping from input to output, with low-dose input itself providing conditional information. In particular, we propose a style-based generator to bridge the discrepancy in style.

\subsection{Style Transfer}	
Style transfer is a transformative task that changes the style of the source image, making its style close to that of the target image while preserving the original content.

Works on style transfer can be divided into two major kinds: (1) constraint-based methods; (2) style modulation methods. Constraint-based methods focus on losses and constraints, like perceptual losses\cite{johnson2016perceptual}, histogram losses\cite{risser2017stable}, and texture networks\cite{ulyanov2016texture}. StarGAN adopts consistent constraints to realize style translation from one domain to multiple domains\cite{choi2018stargan}, which is unsuitable for one-to-one PET translation. StarGAN will degenerate into CycleGAN\cite{supervisedCycleGAN} if applied in one-to-one style transfer bluntly, which also belongs to constraint-based methods.

The other is rooted in style modulation on intermediate features.
Ulyanov \textit{et al.}\cite{ulyanov2016instance} first improve the quality of feed-forward stylization significantly with a modulation module called instance normalization (IN). Dumoulin \textit{et al.}\cite{dumoulin2016learned,ghiasi2017exploring} extend the modulation approach and transfer the input image to an arbitrary user-specified style with conditional instance normalization. Huang \textit{et al.}\cite{huang2017arbitrary} conclude that normalizing feature statistics and performing scaling and shifting, which can be described as style modulation, are the keys to changing style. Huang \textit{et al.} further propose adaptive instance normalization (AdaIN) to realize an arbitrary style transfer in real-time. These approaches do not require complex designs of loss function; it is efficient and compact in modulating textures, achieving state-of-the-art performance in style transfer.

StyleGAN\cite{karras2019style,karras2020analyzing} leverages a style modulation approach from the aforementioned works\cite{ulyanov2016instance,dumoulin2016learned,huang2017arbitrary} and can generalize portraits with promising textures and vivid details. Nonetheless, StyleGAN is typically for random portrait generation so that it is incapable of image-to-image translation because of its uncertainty. SPADE\cite{park2019semantic} realizes an image-to-image translation method from labeled semantic images to natural images with style modulation, but it takes a random vector as the input of the generator, which is inadequate for tasks with stronger constraints, like PET synthesis.

Enlightened by StyleGAN, we introduce style modulation to bridge the style gap between full-dose and low-dose domains. Compared with StyleGAN, our SGSGAN: (1) is an image-to-image translation framework for PET synthesis; (2) is a 3D network structure for PET image synthesis, whereas StyleGAN is a 2D framework and has too many parameters when generalized to 3D; (3) adopts the segmentation guided strategy to further boost the performance of synthesis. Compared with SPADE, SGSGAN strengthens input constraints and generates reliable outputs.

\subsection{Task-driven Strategy}
Task-driven strategy has been prevailing in generative tasks recently\cite{ghafoorian2018gan,yang2019knowledgeable,kim2019low,segmentationGuided,chaitanya2019semi,zhang2020unsupervised,tang2019tuna}. Essentially, task-driven strategy introduces the most needed semantic task-specific constraint into the training process of the generative models, forms an end-to-end structure, and thus boosts the generation performance.

Methods using this strategy vary depending on the tasks. EL-GAN\cite{ghafoorian2018gan} uses an embedding-loss-driven strategy for lane detection. Yang \textit{et al.}\cite{yang2019knowledgeable} use an external commonsense knowledge-driven strategy for visual storytelling. Kim \textit{et al.}\cite{kim2019low} adopt a local-illumination-driven strategy at the training-set level for enhancing low-light images. Jiang \textit{et al.}\cite{segmentationGuided} leverage a segmentation-driven strategy to improve the performance of image translation, alleviating the lack of spatial controllability. 

In the medical image synthesis field, TD-GAN\cite{zhang2020unsupervised} successfully achieves better synthesis and more favorable segmentation simultaneously for unseen real X-ray images and suggests the practicality of task-driven strategy. Tang \textit{et al.}\cite{tang2019tuna} propose TUNA-Net to favor the target disease recognition task and gain a competitive result. 

The aforementioned works prove that task-driven strategy is prone to benefiting 3D PET synthesis in task-related regions. Meanwhile, the segmentation of ROIs in PET is of importance in some applications \cite{hsu2008automatic,ren2019atlas,wong2002segmentation,hatt2009fuzzy,song2013optimal,ju2015random}. Thus, we propose a segmentation guided strategy to enhance the performance of 3D full-dose PET synthesis, especially in segmentation-task-related regions.  

\section{Methods}
\label{sec_method}
\begin{figure}[t]
\centering
\includegraphics[width=\linewidth]{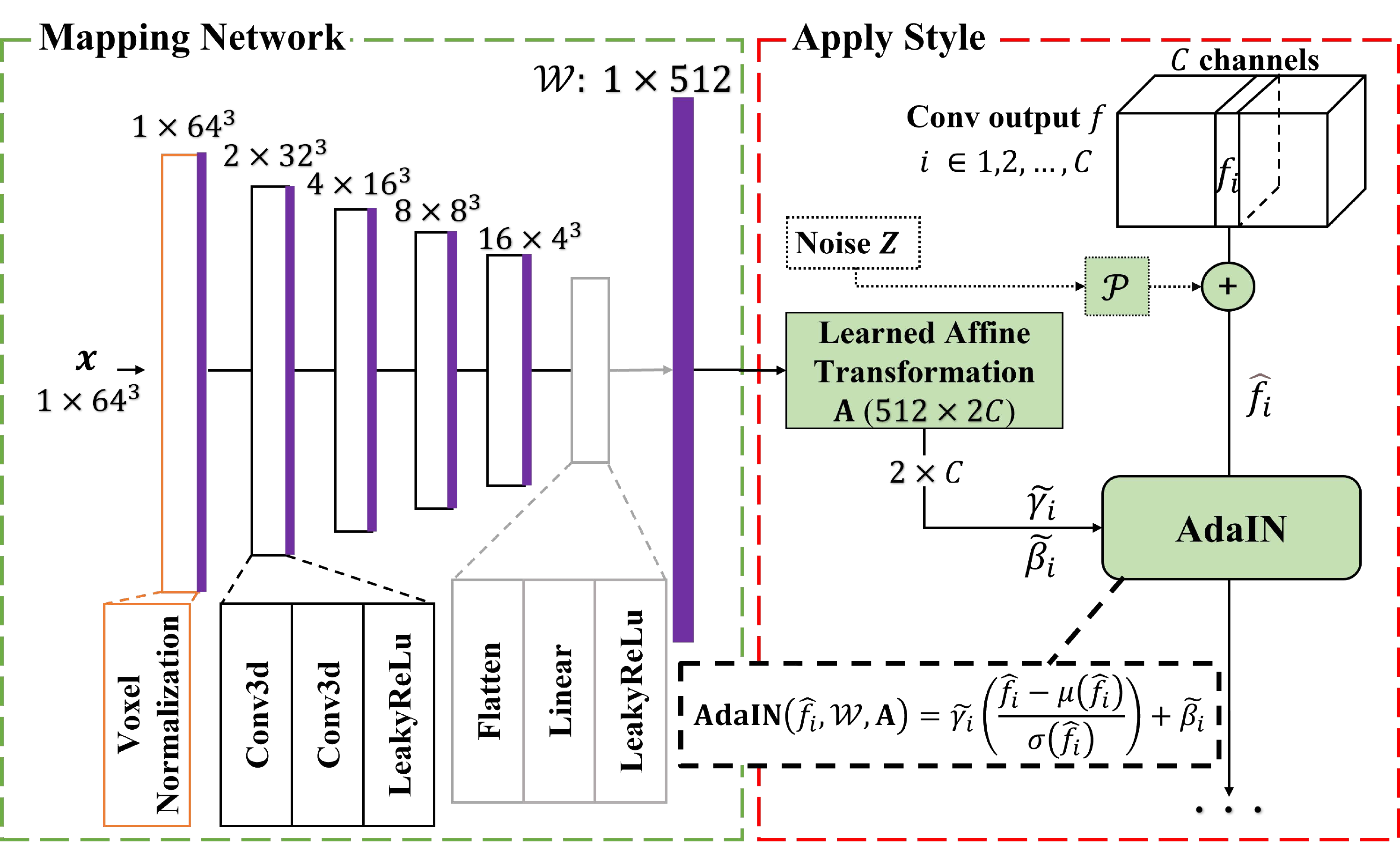}
\caption{Illustration of style modulation. The mapping network (left) takes 3D low-dose PET $x$ as input and generates the style representation $\mathcal{W}$. Style modulation is conducted in different feature layers. Each single-channel feature map $f_{i}$ outputted from the upper convolutional layer is added with noise $Z$ scaled by learned scaling parameter $\mathcal{P}_{i}$. Then the noise-injected feature map $\hat{f}_{i}$ is modulated by style factors $\left\langle\tilde{\gamma}_{i},
\tilde{\beta}_{i}\right\rangle$ produced from $\mathcal{W}$ by layer-specific learned affine transform $\mathbf{A}$. Note that the noise module is enabled in training and disabled in testing, which is marked with black dash lines.}
\label{AdaIN}
\end{figure}

Our goal is to establish a 3D image synthesis model capable of translating low-dose PET to full-dose PET with reliable textures and favorable details, especially in ROIs of clinical concern under the guidance of segmentation.

One training triplet is denoted by $\left(x,y,s\right)$, where $x$ refers to 3D low-dose PET image, $y$ refers to the corresponding paired 3D full-dose PET image, and $s$ is the ROIs segmentation mask of $y$ (details discussed in \ref{section:dataset}). The generated full-dose image is denoted by $\hat{y}$ and the intermediate predicted ROIs mask is denoted by $\hat{s}$.
 
We adopt GAN\cite{goodfellow2014generative} as our basic image synthesis model, which is composed of a generator ($G$) and a discriminator ($D$). The output of the generator is denoted as $G\left(\cdot\right)$ and the output of the discriminator is denoted as $D\left(\cdot \right)$. The segmentation network, which outputs the segmentation mask and imposes the segmentation constraint on $G$, is denoted as $S$ and its output as $S\left(\cdot \right)$. Thus, our goal can be formulated as establishing a mapping $G\left(x\right) \longrightarrow \hat{y}$ under the constraint $S\left(\hat{y}\right) \sim \hat{s}$.

In the following section, we expatiate the style-based generator $G$, the segmentation guided strategy as well as the objective function and our overall framework in detail. 

\subsection{Style-based Generator}
\label{sec:style-based module}

The content of a PET image is anatomical structures like organs, similar to semantic contents in natural images\cite{johnson2016perceptual}. And the style of a PET image means textures and common patterns, which can be controlled by the mean and standard deviation of the image feature maps \cite{huang2017arbitrary}. Texture differences exist between low-dose PET and full-dose PET, which could be regarded as a style gap. So we expect the generator to narrow the style gap. Recently, style modulation derived from style transfer \cite{dumoulin2016learned,ulyanov2016instance,huang2017arbitrary} has been proved to have a significant effect on the image synthesis model with adversarial loss\cite{karras2019style,karras2020analyzing}. Style modulation can effectively realize the transfer on textures. It can change the feature statistics of the input, furnishing the generated image with higher reliability and fidelity in style. Thus, we design a style-based generator with style modulation for 3D PET synthesis.

The style-based generator consists of a mapping network, a noise module, an adaptive instance normalization (AdaIN) operation, and a synthesis network.

\subsubsection{Mapping Network}

The mapping network generates the style representation from the input for further modulation. The mapping network of StyleGAN\cite{karras2019style} takes a random latent vector as the input and maps it to a style representation. This design results in an uncertainty of generated images and unsuitability for image-to-image translation. 

Different from StyleGAN, our mapping network takes a certain low-dose image as input instead of a random latent vector. By doing so, we can ensure the style representation performing subsequent modulation is input-specific. The mapping network learns to generate the style representation $\mathcal{W}$ that is most relevant to textures when the generator attempts to bridge the style gap between full-dose and low-dose PET. This more linear, less entangled $\mathcal{W}$ will be utilized to scale and shift different convolutional layer features to adjust the style on the feature level. 

The architecture of the mapping network is shown in \autoref{AdaIN}. The input is normalized first. There are four convolution blocks and one linear layer, all of which are followed by a LeakyReLU. Each convolution block consists of two convolution layers: one is for doubling channels with the $2\times2\times2$ filter, the other is for halving the size with the $3\times3\times3$ filter. Finally, the feature maps are flattened and we get the style representation $\mathcal{W}$ at the size of 512, for the reason that this data form is in accordance with the wanted linear space of style\cite{karras2019style}.

\subsubsection{Noise Module}
Adding noise to the training process of convolutional networks can help $G$ jump out of local optimum and improve overall robustness\cite{addnoise}. In our style-based generator, different Gaussian noises are injected after convolutional layers by learned scaling parameters. To be specific, a single-channel random noise
$Z$ conforming to the normal distribution, which is of the same size as the feature map, is scaled using channel-specific learned scaling 
parameter $\mathcal{P}$ and is broadcasted to all channels:
\begin{equation} \hat{f}_{i} = f_{i} + \mathcal{P}_{i}Z, \quad i = 1,2,…C\label{eq1} 
\end{equation}
where $C$ denotes the total number of channels of the last convolution outputs, $f_{i}$ denotes the $i^{th}$ feature map,
$\hat{f}_{i}$ denotes the $i^{th}$ feature injected noise, and $\mathcal{P}_{i}$ denotes the $i^{th}$ scaling parameter.

As long as noise $Z$ consistently conforms to the normal distribution in the training process, the independently injected noise will not affect the unbiasedness of the estimation. We enable the noise module in training and disable it in testing just to simply ensure that the testing output always stays still.

\subsubsection{Adaptive Instance Normalization (AdaIN)}
Instance normalization (IN) is the core structure of style 
transfer, which shifts and scales the input feature statistics to a certain style. Its basis can be formulated as:
\begin{equation}\operatorname{IN}(\hat{f}_{i})=\gamma\left(\frac{\hat{f}_{i}-\mu(\hat{f}_{i})}{\sigma(\hat{f}_{i})}\right)+\beta, \label{eq2}\end{equation}
where $\mu(\hat{f}_{i})$ and $\sigma(\hat{f}_{i})$ denote the mean and the standard deviation of $\hat{f}_{i}$, computed independently for each instance. $\gamma$ and $\beta$ denote style factors, also called scaling and shifting parameters, respectively.

In our work, style factors are learned from the style representation $\mathcal{W}$. $\mathcal{W}$ is a one-dimensional vector, a high-level semantic representation generated by the mapping network, and is highly related to texture. The most direct method to generate different style factors from $\mathcal{W}$ is to use affine transform $\mathbf{A}$. The dimentions of $\mathbf{A}$ are $512 \times 2C$, where $C$ denotes the total number of channels of the corresponding convolution outputs. For distinguishing from IN, our style factors are denoted as $\left\langle\tilde{\gamma}_{i}, \tilde{\beta}_{i}\right\rangle$. The affine transform can be formulated as:
\begin{equation}\left\langle \tilde{\gamma}_{i},\tilde{\beta}_{i} \right\rangle=\mathbf{A}(\mathcal{W}), \label{eq3} \end{equation}

Our AdaIN is slightly different from the AdaIN in \cite{huang2017arbitrary}, whose $\gamma$ and $\beta$ are the standard deviation and the mean of the style image. Due to its adaptive adjustment, we still call the normalization operation in our approach adaptive instance normalization (AdaIN), and it is defined as:
\begin{equation}\operatorname{AdaIN}(\hat{f}_{i},\mathcal{W},\mathbf{A})=\tilde{\gamma}_{i}\left(\frac{\hat{f}_{i}-\mu(\hat{f}_{i})}{\sigma(\hat{f}_{i})}\right)+\tilde{\beta}_{i}. \label{eq4}\end{equation}

To be specific, $\tilde{\gamma}$ and $\tilde{\beta}$ are different for each instance, so if the size of input feature maps $\hat{f}$ is $ N\times C $ (N instances, C channels), the size of $\tilde{\gamma}$ and $\tilde{\beta}$ would be $N\times C$. Different convolutional layers are followed by AdaIN with different learned affine transforms $\mathbf{A}$. AdaIN is active both in training and testing. It makes inputted PET images less entangled when passed through the encoder, thus realizing a style conversion of features on different convolutional layers and making texture translation more effective from low-dose $x$ to full-dose $y$.

\subsubsection{Synthesis Network}
\label{subsubsec:synthesis network}
The synthesis network is the backbone of $G$ for PET synthesis. In our SGSGAN, the synthesis network could be implemented by varied existing backbone networks because the style modulation part is easily embeddable. To achieve better performance and cut down parameters, we design a new 3D synthesis network architecture called densely connected image translation network (DCITN). 

DCITN applies dense connection derived from DenseNet\cite{huang2017densely}. Dense connection has been demonstrated to be effective on MRI super resolution by DCSRN\cite{chen2018brain}. Dense connection forms direct skip connection paths in the network structure. It transmits feature maps from any layer to all subsequent layers and concatenates them by channel dimension. These dense cross-level feature transmission paths reduce the back-propagating burden of the model. Meanwhile, since the feature maps are reused among levels, the amount of parameters is reduced, and this characteristic to some extent makes the whole system resistible to overfit. Furthermore, these paths are capable of grasping global contextual information of PET images efficiently and ensuring fast convergence. 

\begin{figure}[t]
\centerline{\includegraphics[width=\columnwidth]{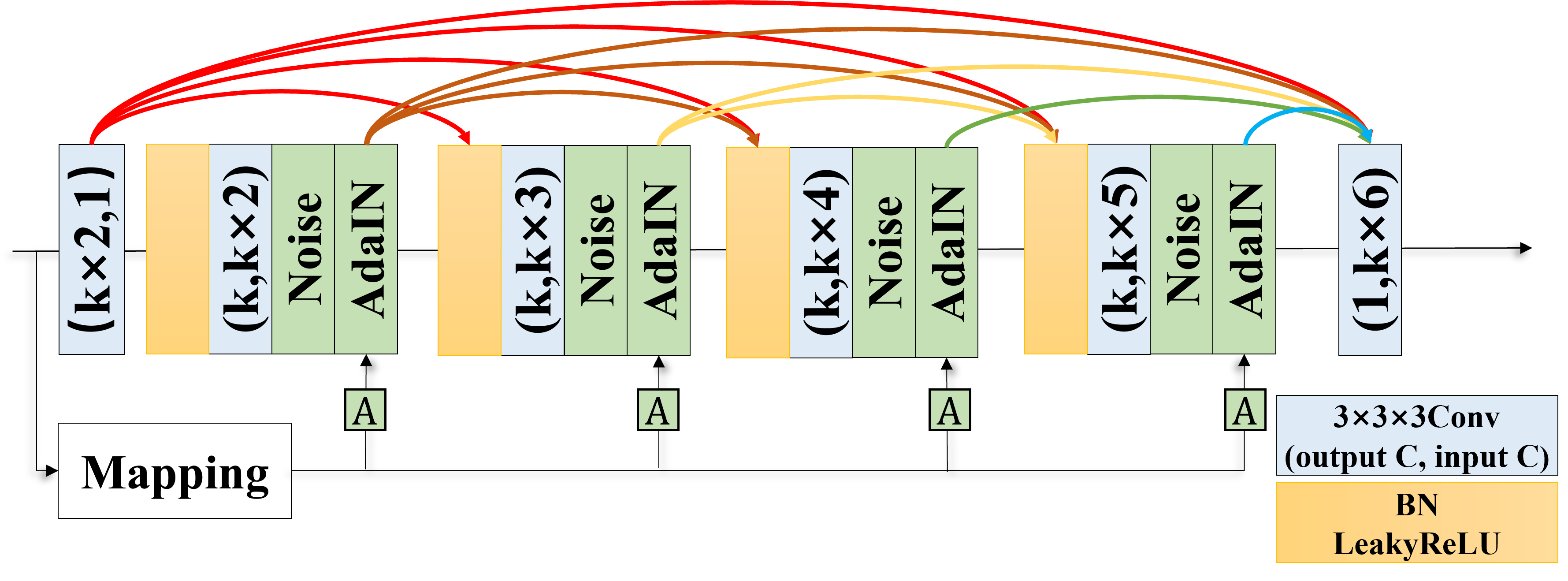}}
\caption{The architecture of the style-based generator with DCITN. DCITN is a choice for the synthesis network in \autoref{method}. It applies dense connection derived from DenseNet\cite{huang2017densely} and is embedded with style modulation. The input channel number and the output channel number are illustrated inside the convolutional layers (blue blocks). $k=8$ in this paper. Note that the noise module is enabled in training and disabled in testing.}
\label{SGSGAN-DCITN}
\end{figure}

The architecture of our DCITN is shown in \autoref{SGSGAN-DCITN}, as a part of the style-based generator. All convolutional layers in DCITN use kernel size of $3\times3\times3$, padding of $1$, and stride of $1$.

\subsection{Segmentation Guided Strategy (SG)}

Clinical experts expect to obtain legible full-dose PET images with high quality in ROIs. Recently, task-driven strategy has been proved effective in various generative tasks\cite{zhang2020unsupervised,ghafoorian2018gan,kim2019low,chaitanya2019semi}. There are two main advantages of task-driven strategy. First, it leverages the most needed information for a specific task, introduces task-specific knowledge into the training process of the generative model, and eventually enhances the effect of generation in accordance with the task. Second, task-driven strategy is easily adjustable because the auxiliary constraint can alter cleverly according to the target task. Specifically, the segmentation task focuses on ROIs of concern, coinciding with PET image translation. Therefore, we propose the segmentation guided strategy (SG) for the problem. 

Our proposed segmentation guided architecture consists of two parts: the main body GAN and the segmentation network $S$. GAN synthesizes full-dose PET $\hat{y}$ from low-dose PET $x$, while $S$ contributes to generating predicted mask $\hat{s}$ from $\hat{y}$ and imposes a segmentation constraint on GAN for further synthesis refinement. The segmentation guided architecture enhances the synthesis performance in ROIs of concern, strongly meeting the need for clinical diagnosis. (ablation study discussed in \ref{sec:ablation seg}).

As is shown in \autoref{method}, $\hat{y}$ is passed to $S$ for predicted ROIs mask $\hat{s}$, with which ground truth mask $s$ compares for a segmentation loss $\mathcal{L}_{Seg}$. $\mathcal{L}_{Seg}$ is then imposed on GAN so that the synthesis performance of segmented ROIs is improved. In the interaction between GAN and $S$, $\mathcal{L}_{Seg}$ might introduce additional segmentation error to GAN, adversely affecting the overall synthesis result. But the number of ``the segmentation error'' accounts for only a small percentage compared with ``the segmentation precision''. From the perspective of machine learning, a large percentage of ``the segmentation precision'' will lead to the performance improvement of synthesis, whereas a small percentage of ``the segmentation error'' will hardly affect the performance of synthesis.

Moreover, our segmentation guided strategy belongs to multi-task learning (MTL). MTL utilizes more data from different learning tasks when compared with single-task learning\cite{zhang2021survey} and thus can learn more robust and universal representations for multiple tasks and more powerful models, leading to better knowledge sharing among tasks, better performance of each task, and low risk of overfitting in each task. The same data is used in two different tasks, which is equivalent to two copies of data being used. Thus, the synthesis enjoys the benefit of MTL and is further enhanced.

In our model, we combine dice loss ($\mathcal{L}_{Dice}$) proposed in \cite{milletari2016vnet} with vanilla BCE loss ($\mathcal{L}_{BCE}$) for a better segmentation network, which is defined as:
\begin{equation}
\begin{aligned}
\mathcal{L}_{BCE} = &  \sum^{all\;voxels} [ s \log \left( \hat{s} \right) \\
&+\left( 1-s \right) \log 
\left( 1-\hat{s} \right)  ] 
\end{aligned} ,
\label{bceloss} 
\end{equation}
\begin{equation}\mathcal{L}_{Dice}=1-\frac{2|\hat{s}\cap
s|}{|\hat{s}|+|s|+\epsilon} ,\label{Diceloss}\end{equation}
\begin{equation}
\mathcal{L}_{Seg} = \mathcal{L}_{BCE}+\mathcal{L}_{Dice} .
\label{lossseg}
\end{equation}
For simplicity, we still use $\hat{s}$ to denote the probability score of the predicted voxel calculated from sigmoid, and $s$ to denote the ground truth binary voxel label in \eqref{bceloss}. $\epsilon$ refers to the smooth parameter.

We chose the most commonly used Unet\cite{2015unet,3d-seg} as the segmentation network $S$. The characteristic of skip-connections in Unet helps it bridge the gradient vanishing when training deep networks and makes it efficient at generating medical masks.

\subsection{Objective Function and Overall Framework}
Our overall framework is shown in \autoref{method}. SGSGAN consists of two parts: the main body GAN, i.e. the style-based generator $G$ and the discriminator $D$, and the segmentation network $S$. GAN is for synthesis, in which $G$ takes the low-dose PET image as input to perform image translation with style modulation and $D$ performs domain discrimination for adversarial learning. $S$ provides an extra constraint for further improvement in the ROIs of concern.

The original generative adversarial network (GAN)\cite{goodfellow2014generative} is proposed for unconditional 2D image generation. Then conditional GAN (cGAN) introduces conditional constraints into GAN to realize some degree of controllability on synthesized images. The representative research among cGAN in paired-images translation is pix2pix\cite{Isola2017image}. To capture the spatial information of 3D images, we implement the pix2pix framework to 3D form. Pix2pix uses Jensen-Shannon divergence to compare the data distributions, resulting in problems such as convergence difficulty and mode collapse. These problems can be bridged to some extent by comparing the generated image data distributions with Wasserstein distance\cite{arjovsky2017wasserstein}.  

We tune the 3D pix2pix framework and adopt Wasserstein distance. The adversarial loss $\mathcal{L}_{Adv}$ of our SGSGAN is defined as:
\begin{equation}
\begin{aligned}
\mathcal{L}_{Adv}=&-\mathbb{E}_{y \sim p_{\text {data}}(y)} \left[ (D(y))\right] \\ &+ \mathbb{E}_{x \sim p_{\text {data}}(x)}\left[(D(G(x)))\right],
\end{aligned} \label{eq_ouradv}
\end{equation}
where $\mathbb{E}$ refers to the maximum likelihood estimation. Two terms in \eqref{eq_ouradv} perform a Wasserstein distance estimation, which removes the last sigmoid layer in $D$ and the log function in the losses. After each gradient update, the weights of $D$ are clipped to $\left[-c,c\right]$ to enforce a Lipschitz constraint, where $c$ is the clipping parameter. $\mathcal{L}_{Adv}$ stands for the adversarial loss, forcing $G$ to synthesize more realistic images which could fool $D$ and forcing $D$ to discriminate the true from the fake. Content loss $\mathcal{L}_{Content}$ is to help $G$ minimize the voxel-wise intensity gap between $G(x)$ and $y$. Without loss of generality, we use a L1 norm as the content loss following \cite{Isola2017image,supervisedCycleGAN,wang20183d}:
\begin{equation}\mathcal{L}_{Content}= \mathbb{E}_{x,y \sim p_{data}(x,y)}\left[\|y-G(x)\|_{1}\right], \label{eq_losscon} \end{equation}
where $\|\cdot\|_{1}$ is a L1 norm. $\mathcal{L}_{Adv}$ and $\mathcal{L}_{Content}$ compose the basic synthesis loss. In addition, $\mathcal{L}_{Seg}$ provides the segmentation guided term, which is produced by $S$. The final objective function of SGSGAN is:
\begin{equation}
\begin{aligned}
\mathcal{L}_{SGSGAN} = &\mathcal{L}_{Adv}+\lambda_{1}\mathcal{L}_{Content}+ \lambda_{2}\mathcal{L}_{Seg},
\end{aligned}
\label{eq_sgsgan}\end{equation}
where $\lambda_{1}$ and $\lambda_{2}$ are hyperparameters to balance $\mathcal{L}_{Adv}$, $\mathcal{L}_{Content}$ and $\mathcal{L}_{Seg}$.

The generator $G$ has been discussed in \ref{sec:style-based module} (see \autoref{SGSGAN-DCITN}). Besides our DCITN, the synthesis network could be implemented by other network architectures, like Unet\cite{modified-Unet}.

Our $D$ adopts the main structure of patchGAN from pix2pix\cite{Isola2017image}. Different from pix2pix, the input is not concatenated with the original image $x$ to reduce the calculation. The sigmoid layer is removed from $D$ to perform a Wasserstein distance estimation.

Together with the segmentation network $S$, it is our proposed SGSGAN shown in \autoref{method}. In the actual training process, we take the strategy of alternate training between GAN and $S$. 

In this process, $G$ and $D$ are trained simultaneously in one iteration until reaching an adversarial balance, whereas $S$ is trained independently until convergence. The training processes of the main body GAN and $S$ are detached but alternating. The alternate training process can be divided into four phases: (1) Only train GAN until convergence. This phase minimizes $\mathcal{L}_{SGSGAN} = \mathcal{L}_{Adv}+\lambda_{1}\mathcal{L}_{Content}$, with setting $\lambda_{2}$ to 0. (2) Freeze GAN. Train $S$ with the generated image $\hat{y}$ from the frozen $G$ as inputs and minimize the objective $\mathcal{L}_{SGSGAN} = \mathcal{L}_{Seg}$. (3) Freeze the gradient of $S$ and train GAN again with the objective $\mathcal{L}_{SGSGAN} = \mathcal{L}_{Adv} + \lambda_{1}\mathcal{L}_{Content}+ \lambda_{2} \mathcal{L}_{Seg}$, until reaching a new adversarial balance. (4) Restart the training of $S$ with the generated images from the frozen $G$ and the objective $\mathcal{L}_{SGSGAN} = \mathcal{L}_{Seg}$. The training details are discussed in \ref{subsec:Implementation}.

\section{Experiments}

\subsection{Dataset}
\label{section:dataset}

\subsubsection{PET Images}
We evaluate our proposed methods on a full-body PET dataset including 105 patients, which were collected in Peking Union Medical College Hospital. Full-dose PET scans are acquired on the PoleStar m660 PET/CT system with an injection dose of around 0.1$mCi/kg$ of 18F-fluorodeoxyglucose (FDG). The low-dose raw data is synthesized at dose reduction factor DRF = 6 by randomly selecting $1/6$ of the raw count list-mode datasets following the method described in \cite{kim2018penalized}. Then both low-dose and full-dose PET images are reconstructed using standard OSEM methods\cite{OSEM}. For each patient, we sample two low-dose PET images from a single full-dose one, which is an approach for data augmentation to avoid overfitting\cite{xie2019generative}. Thus, we have a total of 210 pairs of (low-dose image, full-dose image) for experiments.
\subsubsection{Selection of ROIs}
To prove the performance of segmentation guidance in preserving specific regions, we select four organs - liver, brain, kidney, and bladder. They are significative for reflecting the quality of PET images \cite{wahl2009recist,boellaard2010fdg}. The liver is characterized by being large and uniform and is often used to assess the uniformity and the noise level of the image intuitively. The brain is chosen for the reason that the clear cortices in the gray matter help to directly evaluate the level of image spatial resolution and noise. The kidney is also used to evaluate the recovery of image contrast and the sharpness of the organ boundary. The bladder can be used to evaluate the quantitative accuracy of SUV in large dynamic range for that the uptake value in this area is generally high.
\subsubsection{Annotations}
\label{sec_annotation} 
Voxel-level annotations of ROIs are provided by three experienced doctors. We follow the annotation strategy in \cite{Jia_2017}: (1) if the labels from two doctors overlap more than $80\%$, we take the intersection, (2) else a senior doctor will step in to give the final annotation. So in total, we have 210 image tuples of (low-dose image $x$, full-dose image $y$, segmented PET mask $s$) from 105 patients, the original size of which is $192\times192\times405$. Without loss of generality, all PET images are resized into $64\times64\times64$ as input using bicubic interpolation to simply investigate how the style-based generator and SG will influence the synthesis performance. We conduct a 21-fold cross-validation on 105 patients to evaluate the model performance, in which 100 patients (200 image tuples) are for training and 5 patients (10 image tuples) are for testing. There is no intersection between patients for training and patients for testing in each experiment. The numerical results in the following tables present the average performance on testing data in 21-fold cross-validation.
\subsubsection{Simulated PET Images}
We also use a set of simulated data for model performance assessment, because it has known ground truths so that we can easily conduct an external validation study on it. The simulated data, XCAT phantom, is constructed following \cite{4DXCAT2010}. The XCAT phantom realistically models the complex shapes of real human organs and simulates a 3D distribution of emission radionuclide activity for the various organs. The data are finally reconstructed with the maximum likelihood expectation maximization (MLEM) algorithm. We generate 200 pairs of simulated data, the original size of which is $192 \times 192 \times 405$. All simulated data are resized into $64\times64\times64$ and used to assess the synthesis performance of our proposed methods.

\subsection{Evaluation Metircs}
PET images aligned with CT contain both anatomical information and functional information. To assess the image recovery performance, two kinds of evaluation method are involved: (1) Math metrics such as peak signal-to-noise ratio (PSNR), structure similarity index (SSIM), and mean absolute error (MAE) are calculated to directly evaluate the quality of the generated images, where the large value of the first two and the small value of the last one imply clear anatomical structure and effective translation performance. (2) Following Isola's work \cite{Isola2017image}, we employ a Unet trained on full-dose image and mask pairs ($y$, $s$) as a special way to measure the discriminability of ROIs in the generated images. Generally, the voxel classification results from the fixed pre-trained Unet model will largely depend on the quality of input PET images. If we synthesize more realistic PET images, we shall get more accurate segmentation results from the pre-trained Unet. We choose the dice coefficient to calculate the voxel classification result and call it Unet-score. It serves not as a specific task-based evaluation metric for the segmentation task, but a general assessment for the quality of synthesized PET images. Moreover, for our methods with SG, we can get another voxel classifier, the segmentation network $S$ used as guidance. We call the corresponding dice value Unet-score-SG. All metrics are calculated to evaluate both the whole body and ROIs. We calculated the metrics within organs by multiplying the images by the corresponding mask and comparing a single organ.

\subsection{Implementation}
\label{subsec:Implementation}
All experiments are implemented on Pytorch using a workstation with NVIDIA V100 GPU (32GB memory). The training procedure and details are described as follows.
\subsubsection{Alternate Training}
For our methods using the segmentation guided strategy (SG), a segmentation task is involved to serve as guidance to preserve ROIs. The segmentation network $S$ used as guidance is trained with ($G(x)$, $s$). $\mathcal{L}_{Seg}$ will contribute to updating $S$ or back-propagating to guide the generator $G$, according to the specific round setting. We conduct an alternate training strategy that fixes one network when training another one. We first train GAN for 1200 epochs, next $S$ for 100 epochs, then GAN for 300 epochs, and finally $S$ for 100 epochs. One looping of this training process is chosen after cross-validation. Conducting more looping changes little to the synthesis performance.

\subsubsection{Noise Setting}
The injecting noise is normally distributed with a standard deviation of 0.01. We apply noise in the training process and disable it in testing.

\subsubsection{Hyperparameter}
All deep-learning-based experiments exploit WGAN to make better use of the model expression ability. The non-momentum-based optimizer RMSprop is chosen to optimize the parameters with a fixed learning rate of $\alpha = 1.0\times10^{-4}$. Weight clipping parameter $c$ is set to 0.01 to enforce Lipschitz constraints. Weighting parameters $\lambda_{1}$ and $\lambda_{2}$ are empirically set to 100 and 1, respectively, to balance the tradeoff among $L_{
Adv}$, $L_{Content}$ and $L_{Seg}$.

\subsection{Comparison}

\begin{table*}[]
\caption{\label{tab:cross_val}
Performance Comparison of Various Methods. Par, L, Br, K, and Bl are the abbreviations for Parameters, Liver, Brain, Kidney, and Bladder, respectively. For PSNR, SSIM, and MAE, ``All'' indicates the performance on the whole image. For Unet-score, ``All'' indicates the result on four organs. * denotes Unet-score-SG}
\resizebox{\textwidth}{!}{%
\begin{tabular}{c|c|ccccc|ccccc|ccccc|ccccc}
\toprule
\multirow{2}{*}{\textbf{Method}} & \multirow{2}{*}{\textbf{\#Par(M)}} & \multicolumn{5}{c|}{\textbf{PSNR↑}} & \multicolumn{5}{c|}{\textbf{SSIM↑}} & \multicolumn{5}{c|}{\textbf{MAE↓}} & \multicolumn{5}{c}{\textbf{Unet-score↑}} \\ \cline{3-22} 
 &  & \textbf{L} & \textbf{Br} & \textbf{K} & \textbf{Bl} & \textbf{All} & \textbf{L} & \textbf{Br} & \textbf{K} & \textbf{Bl} & \textbf{All} & \textbf{L} & \textbf{Br} & \textbf{K} & \textbf{Bl} & \textbf{All} & \textbf{L} & \textbf{Br} & \textbf{K} & \textbf{Bl} & \textbf{All} \\ \hline
Full-seg & \textbackslash{} & \textbackslash{} & \textbackslash{} & \textbackslash{} & \textbackslash{} & \textbackslash{} & \textbackslash{} & \textbackslash{} & \textbackslash{} & \textbackslash{} & \textbackslash{} & \textbackslash{} & \textbackslash{} & \textbackslash{} & \textbackslash{} & \textbackslash{} & 0.9411 & 0.9432 &	0.9422 & 0.9418 & 0.9450 \\
Low-dose & \textbackslash{} & 25.84 & 21.92 & 21.68 & 24.71 & 25.69 & 0.8133 & 0.8371 & 0.8112 & 0.7829 & 0.7047 & 136.87 & 87.56 & 101.34 & 97.75 & 205.65 & 0.8087 & 0.8065 & 0.8095 & 0.8064 & 0.8083 \\
BM3D\cite{bm3d} & \textbackslash{} & 29.19 & 23.67 & 22.32 & 25.89 & 28.88 & 0.9342 & 0.9090 & 0.9011 & 0.9254 & 0.7051 & 51.22 & 20.31 & 38.27 & 49.13 & 124.19 & 0.8727 & 0.8711 & 0.8733 & 0.8730 & 0.8726 \\
RED-CNN\cite{red-cnn} & 1.85 & 31.67 & 24.99 & 20.08 & 29.21 & 32.04 & 0.8821 & 0.8589 & 0.8997 & 0.9073 & 0.6756 & 41.56 & 14.33 & 29.18 & 28.01 & 74.88 & 0.8759 & 0.8775 & 0.8765 & 0.8764 & 0.8763 \\
3D-cGAN\cite{wang20183d} & 53.62 & 31.74 & 25.08 & 20.31 & 29.372 & 32.19 & 0.9895 & 0.9723 & 0.9752 & 0.9907 & 0.7930 & 39.36 & 12.56 & 27.23 & 27.72 & 74.14 & 0.8787 & 0.8774 & 0.8795 & 0.8771 & 0.8785 \\
CycleWGAN\cite{supervisedCycleGAN} & 115.62 & 29.63 & 24.06 & 22.89 & 26.91 & 29.54 & 0.9342 & 0.9648 & 0.9237 & 0.9601 & 0.7860 & 51.22 & 20.31 & 38.27 & 49.13 & 117.48 & 0.8738 & 0.8722 & 0.8743 & 0.8741 & 0.8737 \\
3D-Unet\cite{modified-Unet} & 49.40 & 29.85 & 24.24 & 23.11 & 27.19 & 30.74 & 0.9843 & 0.9819 & 0.9833 & 0.9837 & 0.7602 & 47.34 & 18.77 & 30.18 & 37.45 & 97.91 & 0.8745 & 0.8763 & 0.8745 & 0.8770 & 0.8750 \\
DCITN & 0.03 & 32.30 & 25.56 & 21.22 & 29.69 & 33.11 & 0.9939 & 0.9974 & 0.9885 & 0.9912 & 0.8125 & 34.30 & 6.73 & 18.03 & 22.02 & 65.15 & 0.8841 & 0.8856 & 0.8846 & 0.8766 & 0.8839 \\ \hline
\multirow{2}{*}{SGSGAN-3D-Unet} & \multirow{2}{*}{50.28} & \multirow{2}{*}{30.51} & \multirow{2}{*}{25.47} & \multirow{2}{*}{25.17} & \multirow{2}{*}{28.64} & \multirow{2}{*}{31.46} & \multirow{2}{*}{0.9929} & \multirow{2}{*}{0.9969} & \multirow{2}{*}{0.9968} & \multirow{2}{*}{0.9876} & \multirow{2}{*}{0.7782} & \multirow{2}{*}{42.74} & \multirow{2}{*}{8.66} & \multirow{2}{*}{18.44} & \multirow{2}{*}{30.07} & \multirow{2}{*}{80.31} & \multirow{2}{*}{0.8776} & \multirow{2}{*}{0.8758} & \multirow{2}{*}{0.8782} & \multirow{2}{*}{0.8779} & 0.8775 \\
 &  &  &  &  &  &  &  &  &  &  &  &  &  &  &  & & & & & & 0.8978*  \\ \cdashline{1-22}[2pt/5pt]
\multirow{2}{*}{SGSGAN-DCITN} & \multirow{2}{*}{0.59} & \multirow{2}{*}{\textbf{32.50}} & \multirow{2}{*}{\textbf{27.05}} & \multirow{2}{*}{\textbf{26.26}} & \multirow{2}{*}{\textbf{31.07}} & \multirow{2}{*}{\textbf{33.54}} & \multirow{2}{*}{\textbf{0.9948}} & \multirow{2}{*}{\textbf{0.9980}} & \multirow{2}{*}{\textbf{0.9978}} & \multirow{2}{*}{\textbf{0.9926}} & \multirow{2}{*}{\textbf{0.8188}} & \multirow{2}{*}{\textbf{31.75}} & \multirow{2}{*}{\textbf{5.91}} & \multirow{2}{*}{\textbf{14.54}} & \multirow{2}{*}{\textbf{20.05}} & \multirow{2}{*}{\textbf{63.17}} & \multirow{2}{*}{\textbf{0.8845}} & \multirow{2}{*}{\textbf{0.8863}} & \multirow{2}{*}{\textbf{0.8858}} & \multirow{2}{*}{\textbf{0.8863}} & \textbf{0.8853} \\
 &  &  &  &  &  &  &  &  &  &  &  &  &  &  &  & & & & & & \textbf{0.9014*} \\ \bottomrule
\end{tabular}%
}
\end{table*}

\begin{figure*}[]
    \centering
	\includegraphics[width=\textwidth]{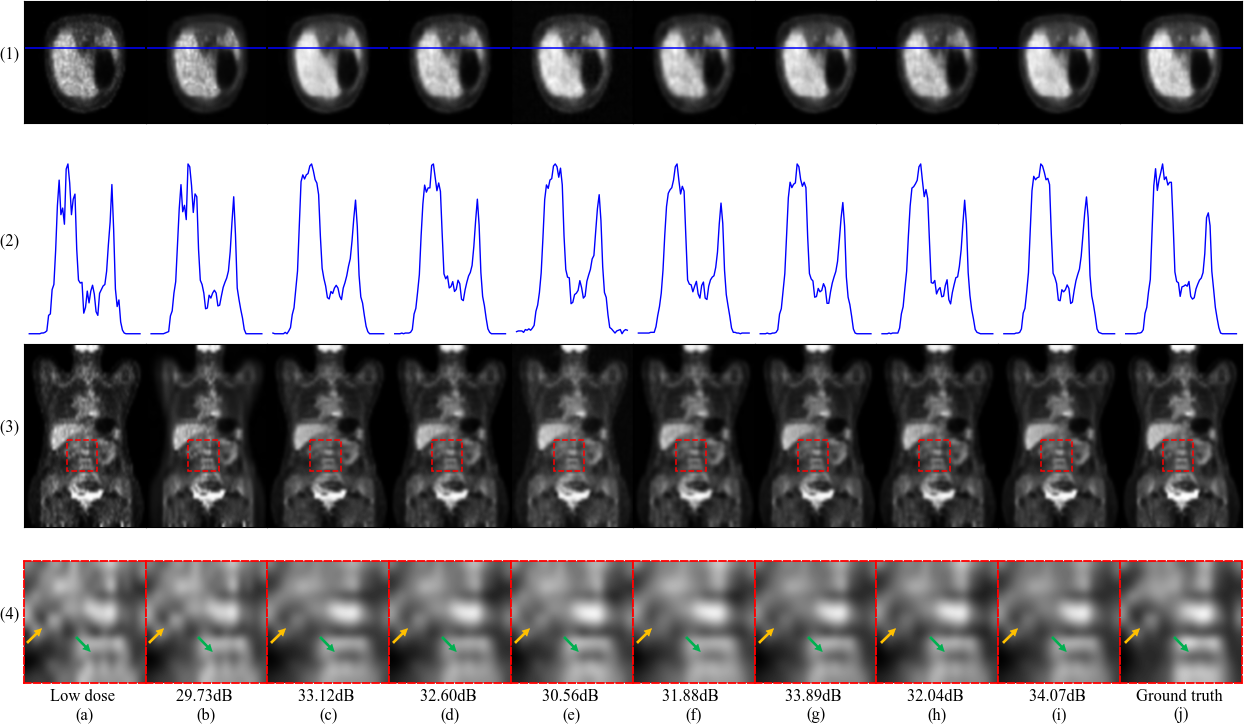}
	\caption{Translation performance comparison of various methods. \textcolor{darkkcyan}{(a)} Low-dose input. \textcolor{darkkcyan}{(b)} BM3D.  \textcolor{darkkcyan}{(c)} RED-CNN. \textcolor{darkkcyan}{(d)} 3D-cGAN.
	\textcolor{darkkcyan}{(e)} 
	CycleWGAN.
	\textcolor{darkkcyan}{(f)} 3D-Unet.
	\textcolor{darkkcyan}{(g)} DCITN. \textcolor{darkkcyan}{(h)} SGSGAN-3D-Unet.  \textcolor{darkkcyan}{(i)} SGSGAN-DCITN. \textcolor{darkkcyan}{(j)} Ground truth. Rows are different selected views: (1) Transverse slices. (2) Plots of the selected line in transverse planes. (3) Coronal slices. (4) Zoomed images of the red rectangles in (3). Arrows point to the regions with obvious differences.}
	\label{fig_method_comparison}
\end{figure*}

\subsubsection{Baselines}
The synthesis network in our proposed SGSGAN could be various architectures. Without loss of generality, we implement two different synthesis models: one is our proposed densely connected image translation network DCITN, the other is 3D-Unet from \cite{modified-Unet}. We denote them as SGSGAN-DCITN and SGSGAN-3D-Unet, respectively.

Our proposed methods are compared with BM3D\cite{bm3d}, 3D-Unet\cite{modified-Unet}, RED-CNN\cite{red-cnn}, 3D-cGAN\cite{wang20183d}, and CycleWGAN\cite{supervisedCycleGAN}. BM3D is an efficient traditional denoising algorithm using sparse representation in the transform domain\cite{bm3d}. 3D-Unet utilizes a modified Unet structure for unsupervised PET denoising in the original project\cite{modified-Unet}. Here we borrow its generator as our comparison model. RED-CNN is proposed for CT denoising\cite{red-cnn} and has also been used for PET translation\cite{supervisedCycleGAN}. 3D-cGAN and CycleWGAN are proposed for PET image synthesis with GAN. Among them, RED-CNN and 3D-cGAN are reported to achieve state-of-the-art performance.

\subsubsection{Results}

The quantitative results from different approaches are summarized in \autoref{tab:cross_val}. The first row is the upper limit of Unet-score of the full-dose image. In the second row, the first 3 columns directly compare low-dose images with full-dose images and the last column shows the Unet-score of low-dose images. Moreover, if the segmentation network for evaluation is directly trained using low-dose images, the overall dice of low-dose images is 0.8384 and the dice within organs are 0.8387, 0.8393, 0.8379, 0.8365, counted on liver, brain, kidney, and bladder, respectively. These results help to verify whether the translation can improve the segmentation task. Since all dice values in \autoref{tab:cross_val} are Unet-score, we do not include the above results in \autoref{tab:cross_val} to avoid confusion.

Apparently, deep-learning-based methods are much more effective than the traditional algorithm BM3D. RED-CNN achieves a satisfactory PSNR result whereas the SSIM is relatively poor. 3D-Unet performs badly compared with 3D-cGAN, which is also a Unet-like architecture. Although having good performance in SSIM and MAE, CycleWGAN is more complex in architecture and does not perform well for PSNR. In contrast, our framework with style modulation and the segmentation guided strategy improves the generation ability of both 3D-Unet and DCITN. In particular, it is worth noting that the image quality enhancement of SGSGAN-3D-Unet and SGSGAN-DCITN on ROIs is even more significant than on the whole image compared with 3D-Unet and DCITN. All these results demonstrate the versatility and the advancement of our SGSGAN.

\begin{figure*}[t]
    \centering
	\includegraphics[width=\textwidth]{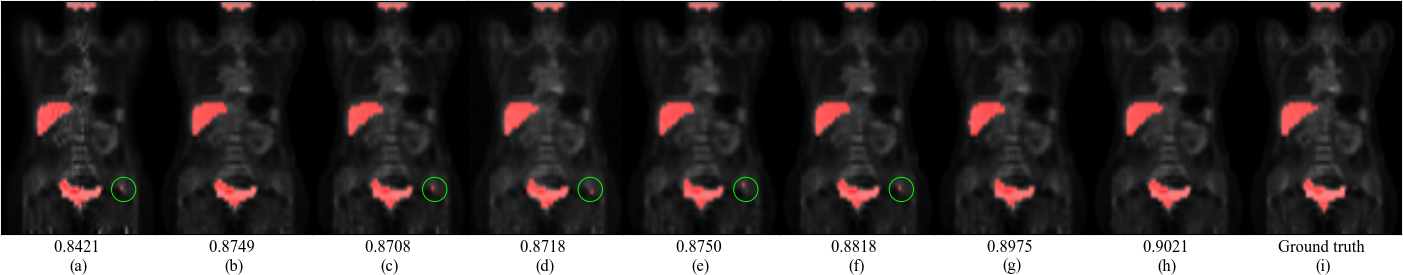}
	\caption{Segmentation performance comparison of various methods. PET images are semi-transparently overlaid by masks. Red regions are masks and green circles point out examples of obvious missegmented areas. \textcolor{darkkcyan}{(a)} BM3D. \textcolor{darkkcyan}{(b)} RED-CNN. \textcolor{darkkcyan}{(c)} 3D-cGAN.
	\textcolor{darkkcyan}{(d)} 
	CycleWGAN
	\textcolor{darkkcyan}{(e)}
	3D-Unet.
	\textcolor{darkkcyan}{(f)} DCITN. \textcolor{darkkcyan}{(g)} SGSGAN-3D-Unet. \textcolor{darkkcyan}{(h)} SGSGAN-DCITN. \textcolor{darkkcyan}{(i)} Ground truth. }
	\label{fig_segmentation_comparison}
\end{figure*}

Example images synthesized from different approaches are shown in \autoref{fig_method_comparison}. Apparent noises still exist in the results of BM3D, 3D-Unet, 3D-cGAN and SGSGAN-3D-Unet (See \autoref{fig_method_comparison}, (1) and (2)). Images produced by 3D-cGAN and 3D-Unet are more blurry, where the anatomy and background are mixed together (See \autoref{fig_method_comparison}, (4)). By contrast, our framework can improve the sharpness of images and SGSGAN-DCITN gives the best visual effect among all.

Generally, the Unet-score increases with the image quality. The Unet-score within organs of SGSGAN-3D-Unet and SGSGAN-DCITN is improved, compared with their counterparts (3D-Unet and DCITN) and other methods. Furthermore, the Unet-score-SG of SGSGAN-3D-Unet and SGSGAN-DCITN are much higher than their Unet-score, which verifies that our framework also conduces to the down-stream segmentation task. \autoref{fig_segmentation_comparison} shows examples of segmentation results. More apparent segmentation errors appear in BM3D,  3D-cGAN, CycleWGAN, 3D-Unet, and DCITN (marked as green circles), whereas SGSGAN-3DUnet and SGSGAN-DCITN well segment the ROIs with SG.

\begin{table*}[t]
\caption{\label{tab:ablation_SMandSG} Ablation studies of Crucial Components. L, Br, K, and Bl are the abbreviations for Liver, Brain, Kidney, and Bladder, respectively. For PSNR, SSIM, and MAE, ``All'' indicates the performance on the whole image. For Unet-score, ``All'' indicates the result on four organs. * denotes Unet-score-SG}
\resizebox{\textwidth}{!}{%
\begin{tabular}{c|cc|ccccc|ccccc|ccccc|ccccc}
\toprule
\multirow{2}{*}{\textbf{Method}} & \multirow{2}{*}{\textbf{Style}} & \multirow{2}{*}{\textbf{SG}} & \multicolumn{5}{c|}{\textbf{PSNR↑}} & \multicolumn{5}{c|}{\textbf{SSIM↑}} & \multicolumn{5}{c|}{\textbf{MAE↓}} & \multicolumn{5}{c}{\textbf{Unet-score↑}} \\ \cline{4-23} 
 &  &  & \textbf{L} & \textbf{Br} & \textbf{K} & \textbf{Bl} & \textbf{All} & \textbf{L} & \textbf{Br} & \textbf{K} & \textbf{Bl} & \textbf{All} & \textbf{L} & \textbf{Br} & \textbf{K} & \textbf{Bl} & \textbf{All} & \textbf{L} & \textbf{Br} & \textbf{K} & \textbf{Bl} & \textbf{All} \\ \hline
3D-Unet &  &  & 29.85 & 24.24 & 23.11 & 27.19 & 30.74 & 0.9843 & 0.9819 & 0.9833 & 0.9837 & 0.7602 & 47.34 & 18.77 & 30.18 & 37.45 & 97.91 & 0.8745 & 0.8763 & 0.8745 & 0.8770 & 0.8750 \\
3D-Unet & \checkmark &  & 30.25 & 25.16 & 25.02 & 28.31 & 31.45 & 0.9926 & 0.9967 & 0.9966 & 0.9865 & 0.7767 & 43.80 & 9.10 & 18.96 & 32.23 & 88.14 & 0.8754 & 0.8731 & 0.8761 & 0.8734 & 0.8750 \\ \cdashline{1-23}[2pt/5pt]
\multirow{2}{*}{3D-Unet} & \multirow{2}{*}{\checkmark} & \multirow{2}{*}{\checkmark} & \multirow{2}{*}{\textbf{30.51}} & \multirow{2}{*}{\textbf{25.47}} & \multirow{2}{*}{\textbf{25.17}} & \multirow{2}{*}{\textbf{28.64}} & \multirow{2}{*}{\textbf{31.46}} & \multirow{2}{*}{\textbf{0.9929}} & \multirow{2}{*}{\textbf{0.9969}} & \multirow{2}{*}{\textbf{0.9968}} & \multirow{2}{*}{\textbf{0.9876}} & \multirow{2}{*}{\textbf{0.7782}} & \multirow{2}{*}{\textbf{42.74}} & \multirow{2}{*}{\textbf{8.66}} & \multirow{2}{*}{\textbf{18.44}} & \multirow{2}{*}{\textbf{30.07}} & \multirow{2}{*}{\textbf{80.31}} & \multirow{2}{*}{\textbf{0.8776}} & \multirow{2}{*}{\textbf{0.8758}} & \multirow{2}{*}{\textbf{0.8782}} & \multirow{2}{*}{\textbf{0.8779}} & \textbf{0.8775} \\
 &  &  &  &  &  &  &  &  &  &  &  &  &  &  &  &  &  &  &  &  &  & \textbf{0.8978*} \\ \hline
DCITN &  &  & 32.30 & 25.56 & 21.22 & 29.69 & 33.11 & 0.9939 & 0.9974 & 0.9885 & 0.9912 & 0.8125 & 34.30 & 6.73 & 18.03 & 22.02 & 65.15 & 0.8841 & 0.8856 & 0.8846 & 0.8766 & 0.8839 \\
DCITN & \checkmark &  & 32.34 & 26.84 & 26.14 & 30.72 & 33.51 & 0.9947 & 0.9979 & 0.9977 & 0.9921 & 0.8187 & 32.38 & 6.18 & 14.77 & 20.91 & 63.41 & 0.8844 & 0.8858 & 0.8857 & 0.8863 & 0.8851 \\ \cdashline{1-23}[2pt/5pt]
\multirow{2}{*}{DCITN} & \multirow{2}{*}{\checkmark} & \multirow{2}{*}{\checkmark} & \multirow{2}{*}{\textbf{32.50}} & \multirow{2}{*}{\textbf{27.05}} & \multirow{2}{*}{\textbf{26.26}} & \multirow{2}{*}{\textbf{31.07}} & \multirow{2}{*}{\textbf{33.54}} & \multirow{2}{*}{\textbf{0.9948}} & \multirow{2}{*}{\textbf{0.9980}} & \multirow{2}{*}{\textbf{0.9978}} & \multirow{2}{*}{\textbf{0.9926}} & \multirow{2}{*}{\textbf{0.8188}} & \multirow{2}{*}{\textbf{31.75}} & \multirow{2}{*}{\textbf{5.91}} & \multirow{2}{*}{\textbf{14.54}} & \multirow{2}{*}{\textbf{20.05}} & \multirow{2}{*}{\textbf{63.17}} & \multirow{2}{*}{\textbf{0.8845}} & \multirow{2}{*}{\textbf{0.8863}} & \multirow{2}{*}{\textbf{0.8858}} & \multirow{2}{*}{\textbf{0.8863}} & \textbf{0.8853} \\
 &  &  &  &  &  &  &  &  &  &  &  &  &  &  &  &  &  &  &  &  &  & \textbf{0.9014*} \\ \bottomrule
\end{tabular}%
}
\end{table*}

\subsection{Ablation studies}
The good generative ability of our method is a result of joint efforts by the synthesis network, style modulation, and the segmentation guided strategy (SG). In this section, we show how they improve the quality of generated images through multiple ablation studies.

\subsubsection{Synthesis Network}
To verify the generalization ability of our framework, we test our framework on two different synthesis networks: 3D-Unet from \cite{modified-Unet} and DCITN. The former is applied because the Unet-like architecture is the most popular architecture in the medical image processing field. Experiments on it would have a reference significance. Experiments demonstrate that the synthesis performance is improved by style modulation and the synthesis quality on those segmentation-task-related regions is enhanced by SG with arbitrary synthesis network architecture. The results indicate that our framework has high generalization ability and the synthesis network could be implemented by varied network architectures. (See \autoref{tab:cross_val}, \autoref{tab:ablation_SMandSG}, and \autoref{tab:ablation_style},  \autoref{fig_method_comparison}, \autoref{fig_segmentation_comparison}, \autoref{fig_styleAblation}).

\subsubsection{Style Modulation and Segmentation Guided Strategy}

The improvement in synthesis compared with the original synthesis network owes to two components, style modulation and SG. The ablation studies of these two crucial components are summarized in \autoref{tab:ablation_SMandSG}. It directly shows that style modulation improves the whole image quality and SG further enhances the quality of task-related regions. To better demonstrate the superiority, we will look into these two components and investigate their effectiveness in detail.

\begin{table}[h]
\centering
\caption{\label{tab:ablation_style}Ablation Studies of Style Modules}
\begin{tabular}{c|cc|ccc}
\toprule
\multicolumn{1}{c|}{\textbf{Method}} & \multicolumn{1}{c}{\textbf{Noise}} & \multicolumn{1}{c|}{\textbf{Adain}} & \textbf{PSNR↑} & \textbf{SSIM↑} & \textbf{MAE↓} \\ \hline
3D-Unet &  &  &30.74& 0.7602 & 97.91 \\
3D-Unet & \checkmark &  &31.08& 0.7708 & 95.33 \\
3D-Unet &  & \checkmark &31.34& 0.7726 & 92.76 \\
3D-Unet & \checkmark & \checkmark &\textbf{31.45}& \textbf{0.7767} & \textbf{88.14} \\ \hline
DCITN &  &  &33.11& 0.8125 & 65.15 \\
DCITN & \checkmark &  &33.35& 0.8161 & 64.89 \\
DCITN &  & \checkmark &33.48& 0.8185 & 64.16 \\
DCITN & \checkmark & \checkmark &\textbf{33.51}& \textbf{0.8187} & \textbf{63.41} \\ \bottomrule
\end{tabular}%
\end{table}

\subsubsection{Style Modulation}
\label{sec:ablation style}

\begin{table*}[h]
\caption{\label{tab:ablation_ROIs}Results of selecting different ROI combinations on SGSGAN-DCITN (mean±standard deviation). The bold indicates the highest performance with the smallest standard deviation. L, Br, K, and Bl are the abbreviations for Liver, Brain, Kidney, and Bladder, respectively. For PSNR, SSIM, and MAE, ``All'' indicates the performance on the whole image. For Unet-score, ``All'' indicates the result on four organs. The paired t-test is conducted between the baseline without SG and an SGSGAN-DCITN with different ROIs as guidance to the significance level of 0.05. When the difference is statistically significant, the corresponding numerical result will be underlined}
\resizebox{\textwidth}{!}{%
\begin{tabular}{c|ccccc|ccccc}
\toprule
\multirow{2}{*}{\textbf{ROIs for SG}} & \multicolumn{5}{c|}{\textbf{PSNR↑}} & \multicolumn{5}{c}{\textbf{SSIM↑}} \\ \cline{2-11} 
 & \textbf{L} & \textbf{Br} & \textbf{K} & \textbf{Bl} & \textbf{All} & \textbf{L} & \textbf{Br} & \textbf{K} & \textbf{Bl} & \textbf{All} \\ \hline
\textbackslash{} & 32.34±1.00 & 26.84±1.22 & 26.14±1.03 & 30.72±1.10 & 33.51±0.89 & 0.9947±0.0034 & 0.9979±0.0021 & 0.9977±0.0018 & 0.9921±0.0029 & 0.8187±0.0091 \\
L & {\ul \textbf{33.23±0.68}} & {\ul 26.79±0.57} & {\ul 26.11±0.53} & {\ul 30.18±0.53} & {\ul 33.52±0.56} & {\ul \textbf{0.9949±0.0018}} & 0.9978±0.0014 & 0.9976±0.0016 & {\ul 0.9922±0.0018} & {\ul 0.8187±0.0027} \\
Br & 32.31±1.08 & {\ul \textbf{27.81±0.80}} & {\ul 26.21±0.58} & 30.73±0.92 & 33.51±0.95 & 0.9947±0.0023 & {\ul \textbf{0.9980±0.0007}} & 0.9977±0.0015 & {\ul 0.9920±0.0019} & {\ul 0.8186±0.0028} \\
K & {\ul 32.41±0.98} & {\ul 27.28±0.66} & {\ul \textbf{26.43±0.77}} & {\ul 30.63±1.05} & 33.52±0.73 & {\ul 0.9948±0.0026} & 0.9979±0.0016 & {\ul \textbf{0.9978±0.0009}} & {\ul 0.9923±0.0024} & {\ul 0.8187±0.0023} \\
Bl & {\ul 32.49±0.60} & {\ul 26.84±0.59} & {\ul 26.04±0.57} & {\ul \textbf{31.32±0.58}} & {\ul 33.52±0.52} & {\ul 0.9947±0.0019} & {\ul 0.9979±0.0012} & 0.9977±0.0014 & {\ul \textbf{0.9926±0.0015}} & {\ul 0.8187±0.0034} \\
L+Br & {\ul 32.48±0.98} & {\ul 27.45±0.79} & 26.13±0.94 & 30.72±1.20 & {\ul 33.53±0.68} & {\ul 0.9948±0.0024} & {\ul 0.9980±0.0011} & 0.9977±0.0021 & 0.9920±0.0027 & {\ul 0.8187±0.0056} \\
L+Br+K & {\ul 32.48±0.98} & {\ul 27.05±1.05} & {\ul 26.37±0.90} & 30.68±0.97 & {\ul 33.53±0.96} & 0.9948±0.0037 & 0.9979±0.0023 & {\ul 0.9978±0.0010} & {\ul 0.9923±0.0032} & {\ul 0.8187±0.0053} \\
L+Br+K+Bl & {\ul 32.50±0.87} & {\ul 27.05±0.84} & {\ul 26.26±0.76} & {\ul 31.07±0.88} & {\ul \textbf{33.54±0.82}} & {\ul 0.9948±0.0029} & {\ul 0.9980±0.0013} & {\ul 0.9978±0.0012} & {\ul 0.9926±0.0027} & {\ul \textbf{0.8188±0.0050}} \\ \bottomrule 
\multirow{2}{*}{\textbf{ROIs for SG}} & \multicolumn{5}{c|}{\textbf{MAE↓}} & \multicolumn{5}{c}{\textbf{Unet-score↓}} \\ \cline{2-11} 
 & \textbf{L} & \textbf{Br} & \textbf{K} & \textbf{Bl} & \textbf{All} & \textbf{L} & \textbf{Br} & \textbf{K} & \textbf{Bl} & \textbf{All} \\ \hline
\textbackslash{} & 32.38±1.39 & 6.18±1.07 & 14.77±0.93 & 20.91±1.28 & 63.41±1.76 & 0.8844±0.0153 & 0.8858±0.0124 & 0.8857±0.0258 & 0.8863±0.0168 & 0.8851±0.0260 \\
L & {\ul \textbf{28.22±0.90}} & {\ul 6.54±1.10} & {\ul 15.93±0.85} & {\ul 20.02±1.22} & {\ul 63.39±1.39} & {\ul \textbf{0.8909±0.0058}} & {\ul 0.8731±0.0157} & 0.8716±0.0357 & {\ul 0.8861±0.0050} & {\ul 0.8831±0.0157} \\
Br & {\ul 30.89±1.53} & {\ul \textbf{5.32±0.98}} & {\ul 14.66±1.13} & {\ul 20.46±1.55} & {\ul 63.40±2.66} & 0.8801±0.0166 & {\ul \textbf{0.8928±0.0057}} & {\ul 0.8873±0.0071} & {\ul 0.8823±0.0261} & {\ul 0.8843±0.0098} \\
K & {\ul 31.91±1.46} & {\ul 5.73±1.17} & {\ul \textbf{13.86±1.12}} & {\ul 20.59±1.12} & {\ul 63.34±1.55} & 0.8807±0.0162 & {\ul 0.8872±0.0083} & {\ul \textbf{0.8972±0.0069}} & {\ul 0.8747±0.0142} & {\ul 0.8847±0.0065} \\
Bl & {\ul 32.04±1.14} & {\ul 5.99±1.23} & {\ul 15.11±0.90} & {\ul \textbf{19.89±1.36}} & {\ul 63.38±1.40} & 0.8850±0.0145 & {\ul 0.8849±0.0055} & {\ul 0.8774±0.0061} & {\ul \textbf{0.8944±0.0055}} & {\ul 0.8844±0.0056} \\
L+Br & {\ul 31.79±1.51} & {\ul 5.91±1.16} & {\ul 14.66±1.28} & {\ul 20.43±1.08} & {\ul 63.35±2.09} & {\ul 0.8852±0.0060} & {\ul 0.8865±0.0061} & {\ul 0.8830±0.0129} & {\ul 0.8830±0.0064} & {\ul 0.8850±0.0062} \\
L+Br+K & {\ul 31.76±2.12} & {\ul 5.93±1.50} & {\ul 14.33±1.33} & {\ul 20.52±1.24} & {\ul 63.26±2.21} & {\ul 0.8849±0.0095} & {\ul 0.8847±0.0102} & {\ul 0.8860±0.0079} & {\ul 0.8837±0.0086} & {\ul 0.8852±0.0084} \\
L+Br+K+Bl & {\ul 31.75±1.78} & {\ul 5.91±1.26} & {\ul 14.54±1.03} & {\ul 20.05±1.50} & {\ul \textbf{63.17±1.72}} & {\ul 0.8845±0.0060} & {\ul 0.8863±0.0068} & {\ul 0.8858±0.0061} & {\ul 0.8863±0.0077} & {\ul \textbf{0.8853±0.0067}} \\ \bottomrule
\end{tabular}
}
\end{table*}

\begin{figure}[t]
\centering
		\includegraphics[width=\columnwidth]{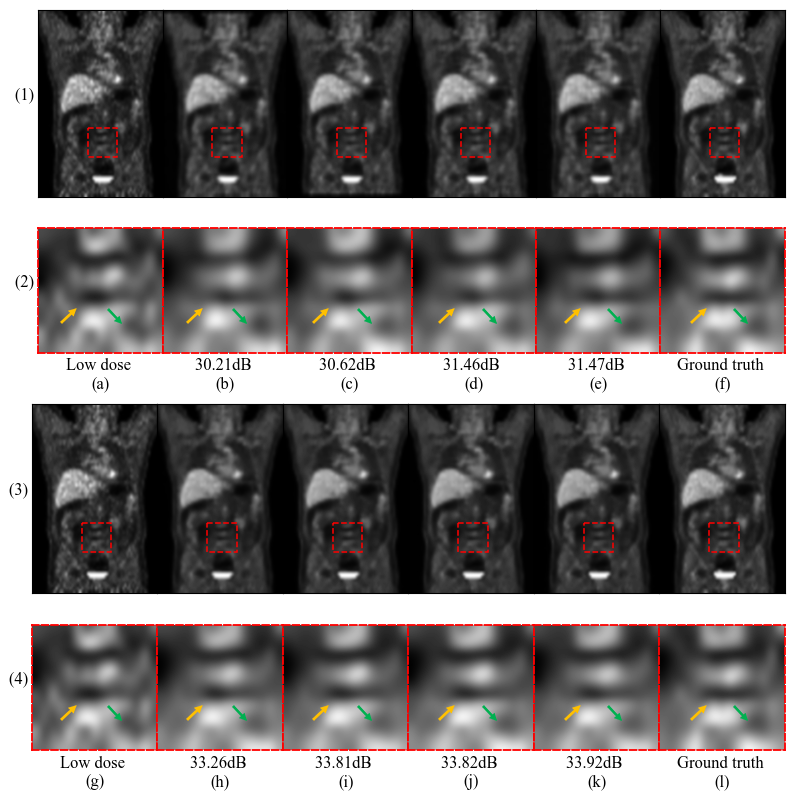}
	\caption{Ablation studies of style modules. The first row of each subfigure displays the coronal plane of the example image while the second row shows the zoomed ROIs in the rectangle region. \textcolor{darkkcyan}{(a)} Low-dose input. \textcolor{darkkcyan}{(b)} 3D-Unet. \textcolor{darkkcyan}{(c)} 3D-Unet+noise. \textcolor{darkkcyan}{(d)} 3D-Unet+AdaIN. \textcolor{darkkcyan}{(e)} 3D-Unet+AdaIN+noise. \textcolor{darkkcyan}{(f)} Ground Truth. \textcolor{darkkcyan}{(g)} Low-dose input. \textcolor{darkkcyan}{(h)} DCITN. \textcolor{darkkcyan}{(i)} DCITN+noise. \textcolor{darkkcyan}{(j)} DCITN+AdaIN. \textcolor{darkkcyan}{(k)} DCITN+AdaIN+noise. \textcolor{darkkcyan}{(l)} Ground truth.}
	\label{fig_styleAblation}
\end{figure}

To verify the performance improvement by style modulation, we apply noise and AdaIN separately to discuss their contributions. \autoref{tab:ablation_style}
displays the results. Both noise and AdaIN can greatly increase the PSNR and the SSIM of both 3D-Unet and DCITN. Applying noise in the training process would give a disturbance to the synthesis network. Although it interferes with capturing features, it could be regarded as an approach of hierarchical feature augmentation, which can thus generate novel feature samples and increase model robustness.

AdaIN achieves more effective performance. It utilizes the learned styles to adjust the features and forces the generator to optimize through style modulation and try to bridge the style gap. Combining noise and AdaIN gives a more promising result. \autoref{fig_styleAblation} shows the examples of using style modules. When adding noise only, the images become sharper and the bumps are more distinct. But some structures are still distorted compared with the ground truth. AdaIN could solve this problem to some extent. Columns (e) and (k) in \autoref{fig_styleAblation}
show that the networks with AdaIN could correct the distorting shape and achieve a better edge similarity. To sum up, applying two modules together leads to the best perceptual quality.

\subsubsection{Segmentation Guided Strategy}
\label{sec:ablation seg}

SG enhances the image synthesis quality on segmentation-task-related regions. To find out how the selection of ROIs affects the SG and the way how SG works, we conduct ablation studies by selecting different ROI combinations on SGSGAN-DCITN. In \autoref{tab:ablation_ROIs}, it turns out that SG does not necessarily improve the overall performance. With SG using different ROIs, the overall performance may fluctuate slightly, but the synthesis performance of each corresponding organ was improved (row 2\textasciitilde5, \autoref{tab:ablation_ROIs}). These results proved that SG has a positive effect on PET synthesis, especially on the segmentation-task-related regions. The overall performance can be improved to some extent, whereas the promotion for a single organ will not be that obvious when involving more and more numbers of organs for SG (row 6\textasciitilde8, \autoref{tab:ablation_ROIs}).  This is a reasonable trade-off because the synthesis performance and the Unet-score within organs still surpass the baseline (row 1, \autoref{tab:ablation_ROIs}).

\subsection{External Validation on Simulated Data}

We use 200 pairs of simulated data, the ground truth, and the simulated low-dose image, to evaluate the effectiveness of style modulation and SG. Here we use the simulated data for the testing of the models trained on real PET data. The comparison results between the generated full-dose images and the ground truths are summarized in \autoref{tab:ablation_simulated}. 

\begin{table}[h]
\centering
\caption{\label{tab:ablation_simulated}Results on simulated data. We use the models trained on real PET data to directly translate the simulated low-dose images into full-dose images}
\begin{tabular}{c|cc|ccc}
\toprule
\textbf{Method} & \textbf{Style} & \textbf{SG} & \textbf{PSNR↑} & \textbf{SSIM↑} & \textbf{MAE↓}\\ \hline
3D-Unet &  &  & 24.17 & 0.6952 & 67.97\\
3D-Unet & \checkmark &  & 24.50 & 0.7184 & 64.88\\
3D-Unet & \checkmark & \checkmark & \textbf{25.11} & \textbf{0.7326} & \textbf{59.43} \\ \hline
DCITN &  &  & 26.09 & 0.7281 & 43.61\\
DCITN & \checkmark &  & 26.26 & 0.7373 & 37.64\\
DCITN & \checkmark & \checkmark & \textbf{27.04} & \textbf{0.7950} & \textbf{34.27}\\ \bottomrule
\end{tabular}%
\end{table}

We also calculate normalized root mean square error (NRMSE), bias, and variance based on SUV between the generated full-dose images and the ground truths. Meanwhile, to better demonstrate the effect of style modulation, we introduce the content loss proposed by \cite{johnson2016perceptual} to measure the content difference (CD) between the generated images and the ground truths. The feature reconstruction loss is the squared normalized Euclidean distance between feature representations coming out from the relu2\_2 of an Imagenet-pretrained VGG-16. We calculate the content difference per slice per axis and get the average. The numerical results are listed in the \autoref{tab:NRMSE}.

\begin{table}[h]
\centering
\caption{\label{tab:NRMSE} NRMSE, Bias, Variance, and CD between the generated images and the ground truth of the simulated data. CD is the abbreviation for Content Difference based on \cite{johnson2016perceptual}}
\begin{tabular}{c|cc|cccc}
\toprule
\textbf{Method} & \textbf{Style} & \textbf{SG} & \textbf{NRMSE↓} & \textbf{Bias↓} & \textbf{Variance↓} & \textbf{CD↓} \\ \hline
3D-Unet &  &  & 0.3467 & 0.7470 & 2.7403 & 0.0050 \\
3D-Unet & \checkmark &  & 0.3455 & 0.7393 &  2.7366 & 0.0046 \\
3D-Unet & \checkmark & \checkmark & \textbf{0.3443} & \textbf{0.7382} & \textbf{2.7328} & \textbf{0.0045}\\ \hline
DCITN &  &  & 0.3323 & 0.4986 & 2.2145 & 0.0042  \\
DCITN & \checkmark &  & 0.3298 & 0.4941 & 2.2007 & 0.0038  \\
DCITN & \checkmark & \checkmark & \textbf{0.3271} & \textbf{0.4938} & \textbf{2.1690} & \textbf{0.0037} \\ \bottomrule
\end{tabular}
\end{table}    

The changing of tendencies of NRMSE, bias, and variance is similar to MAE's, verifying the effectiveness of style modulation and SG. \autoref{tab:NRMSE} also shows that the content difference between the generated images and the ground truths decreases with style modulation, but hardly changes when further applied the segmentation strategy, which means that style modulation plays a more important role in preserving content.

\subsection{Choosing Tumors as ROIs}
\label{sec:choosingtumor}
Preserving tumor shape and contrast is critical in PET denoising. To verify that our method also works when tumors are chosen to be ROIs, 45 patients with tumors are selected from 105 original ones to form a new dataset. We generate 90 corresponding image tuples of (low-dose image, full-dose image, mask of tumors) following the same procedure expatiated in \ref{section:dataset}. We conduct 9-fold cross-validation on this new dataset, where 40 patients are for training and 5 patients are for testing. There is no intersection between patients for training and patients for testing. Here we employ a Unet trained on these full-dose images and their tumor masks to measure the discriminability of these areas in the generated images. The dice coefficient from this Unet is called Unet-score-T. The higher Unet-score-T, the better the tumor detectability. The results are summarized in \autoref{tab:choosing_tumor}. It turns out that the overall synthesized performance is enhanced after style modulation. Furthermore, choosing tumors as ROIs for SG enhances the detectability of corresponding areas, which indicates that SG can be applied to different ROIs.

\begin{table}[h]
\caption{\label{tab:choosing_tumor}Results when choosing tumors as ROIs. For PSNR, SSIM, and MAE, ``All'' indicates the performance on the whole image}
\resizebox{\columnwidth}{!}{
\begin{tabular}{c|c|c|cc|cc|cc|c}
\toprule
\multirow{2}{*}{\textbf{Method}} & \multirow{2}{*}{\textbf{Style}} & \multirow{2}{*}{\textbf{SG}} & \multicolumn{2}{c|}{\textbf{PSNR↑}} & \multicolumn{2}{c|}{\textbf{SSIM↑}} & \multicolumn{2}{c|}{\textbf{MAE↓}} & \textbf{Unet-score-T↑} \\ \cline{4-10} 
 &  &  & \textbf{Tumor} & \textbf{All} & \textbf{Tumor} & \textbf{All} & \textbf{Tumor} & \textbf{All} & \textbf{All} \\ \hline
\multirow{3}{*}{3D-Unet} &  &  & 18.76 & 30.95 & 0.9859 & 0.7613 & 6.74 & 99.79 & 0.9001 \\
 & \checkmark &  & 19.97 & 31.63 & 0.9988 & 0.7778 & 5.31 & 89.63 & 0.9019 \\
 & \checkmark & \checkmark & \textbf{21.53} & \textbf{31.71} & \textbf{0.9997} & \textbf{0.7785} & \textbf{4.03} & \textbf{89.10} & \textbf{0.9106} \\ \hline
\multirow{3}{*}{DCITN} & &  & 21.90 & 32.47 & 0.9917 & 0.7962 & 5.25 & 69.21 & 0.9178 \\
 & \checkmark &  & 22.86 & 33.06 & 0.9995 & 0.8017 & 4.08 & 66.65 & 0.9205 \\
 & \checkmark & \checkmark & \textbf{23.62} & \textbf{33.33} & \textbf{0.9997} & \textbf{0.8125} & \textbf{3.87} & \textbf{66.14} & \textbf{0.9283} \\ \bottomrule
\end{tabular}
}
\end{table}

\section{Discussion}
We propose SGSGAN for generating a full-dose PET image from a low-dose PET with segmentation as guidance and discuss its main characteristics in previous sections. In this section we discuss some tiny factors that could influence the synthesis performance. 

\textbf{Content Loss.} We use a L1 norm as the content loss following \cite{Isola2017image,supervisedCycleGAN,wang20183d} without loss of generality. The choice of content loss is out of scope for our paper but we conducted comparison experiments between the L1 norm, the L2 norm and the gradient difference error on DCITN. Thereinto, we adopt the three-dimensional Sobel operator proposed in \cite{yu2019ea} to calculate the gradient maps and the gradient difference error. The numerical results are shown in \autoref{tab:ablation_contentloss}. The result shows that gradient difference error can't work alone but is a good complement to the L1 norm. Improving the design of image content loss will be our future plan to further enhance the performance.
\begin{table}[h]
\centering
\caption{\label{tab:ablation_contentloss}Comparison Experiments of Content Loss on DCITN}
\begin{tabular}{c|ccc}
\toprule
\textbf{Loss} & \textbf{PSNR↑} & \textbf{SSIM↑} & \textbf{MAE↓} \\ \hline
L1 & 33.11 & 0.8125 & 65.15 \\
L2 & 32.93 & 0.7793 & 68.61 \\
GDE & 26.66 & 0.7023 & 174.84 \\
L1 + GDE & \textbf{33.53} & \textbf{0.8275} & \textbf{64.05} \\ \bottomrule
\end{tabular}%
\end{table}

\textbf{Segmentation Loss.} To construct a better segmentation network, we conduct experiments to choose the segmentation loss. We conduct experiments among dice loss proposed in \cite{milletari2016vnet}, and vanilla BCE loss, and their combination. When using BCE loss and dice loss separately to train the segmentation network with full-dose images, the upper limits of Unet-score are 0.9165 and 0.9389. The combined loss improves the performance to 0.9450. The results are summarized in \autoref{tab:ablation_SegLoss}. The results turn out that the combination works better than using these two losses separately. Therefore, in the training of the task-driven part and the pre-trained voxel classifier for Unet-score, we both adopt the combination loss.
\begin{table}[h]
\centering
\caption{\label{tab:ablation_SegLoss}Ablation Studies of Segmentation Loss}
\begin{tabular}{c|c}
\toprule
\textbf{Loss} & \textbf{Unet-score↑} \\ \hline
BCE & 0.9165 \\
dice & 0.9389 \\
BCE+dice & \textbf{0.9450} \\ \bottomrule
\end{tabular}%
\end{table}

\textbf{More Annotation.} The experiments have demonstrated that the segmentation guided strategy leads to better PET image synthesis, especially on those regions related to the segmentation task. In this paper, these regions are liver, brain, kidney, and bladder. A subset of images with tumors are also tested in \ref{sec:choosingtumor}. Besides these regions, clinicians also care about the synthesis of lesions such as inflammation, infection, and various cancers\cite{foster2014review}. The synthesis performance on such areas will be the future plan after collecting and annotating enough qualified data with lesions.
\par\textbf{Resizing Images.} Global context information has been utilized to benefit the process of capturing style representation as well as reducing false segmentation\cite{GCNet}. We take the whole PET image as input for a global view in this paper. Due to the limitation of GPU memory size, we use bicubic interpolation to resize the original PET images into the size of $64\times64\times64$. This procedure may lead to a change of noise distribution. The synthesis based on patches will be a future plan for this work to explore how the algorithm would perform with images in its native resolution.

\section{Conclusion}
\par In this paper, we propose a novel segmentation guided style-based generative adversarial network (SGSGAN) for translating full-dose PET images from low-dose ones. We design a style-based generator to improve the quality of generated images through style modulation. Besides, the segmentation guided strategy is introduced to improve the translation performance, especially on the segmentation-task-related ROIs. Comprehensive experiments show that our methods achieve state-of-the-art performance in PET translation. In addition, our proposed framework has the potential to be implemented on other networks simply by replacing the backbone of the synthesis network.

Future work will concentrate on the patch-level synthesis, exploring approaches to better utilize style information, and applying it to other image translation tasks in the medical imaging field, like CT, MRI.

\normalem
\bibliographystyle{IEEEtran}
\bibliography{tmi.bib}
\end{document}